\documentstyle[12pt]{article}

\begin{document}
\title{\textbf{Resonant oscillations in ${\alpha}^{2}$-dynamos on a closed, twisted Riemannian 2D flux tubes}} \maketitle
{\sl \textbf{L.C. Garcia de Andrade}\newline
Departamento de F\'{\i}sica
Te\'orica-IF\newline
Universidade do Estado do Rio de Janeiro\\[-1.5mm]
Rua S\~ao Francisco Xavier, 524\\[-1.5mm]
Cep 20550-003, Maracan\~a, Rio de Janeiro, RJ, Brasil\\[-1.5mm]
Electronic mail address: garcia@dft.if.uerj.br\\[-1.5mm]
\vspace{0.01cm} \newline{\bf Abstract} \paragraph*{}Chicone et al [CMP (1995)] have shown that, kinematic fast dynamos in diffusive media, could exist only on a closed, 2D Riemannian manifold of constant negative curvature. This report, shows that their result cannot be extended to oscillatory ${\alpha}^{2}$-dynamos, when there are resonance modes, between toroidal and poloidal frequencies of twisted magnetic flux tubes. Thus, dynamo action can be supported in regions, where Riemannian curvature is positive. For turbulent dynamos, this seems physically reasonable, since recently,  [Shukurov et al PRE (2008)] have obtained a Moebius flow strip in sodium liquid, torus Perm dynamo where curvature is also connected to the magnetic fields via diffusion. This could be done, by adjusting the corresponding frequencies till they achieved resonance. Actually 2D torus, is a manifold of zero mean curvature, where regions of positive and negative curvatures exist. It is shown that, Riemannian solitonic surface, endowed with a steady ${\alpha}^{2}$-dynamo from magnetic filamentary structures [Wilkin et al,PRL (2007)].  \newpage
\section{Introduction}
Turbulent flows in curved configurations, has been a topic of growing interest, among  physicists and mathematicians. In particular, turbulent dynamos \cite{1} in the plasma universe, has been a topic, which have help us, to better understand mechanisms, by which, magnetic fields grow or decay, as in stellar objects \cite{2}. The first important mathematical investigation on kinematic magnetic dynamo has been done by Arnold et al \cite{3} in a flow with uniform stretching in Riemannian space of torus shape. More recent the Perm russian dynamo torus group \cite{4,5} have designed and built an experiment with turbulent dynamos, by performing a breaking in rotating two dimensional metalic torus filled with liquid sodium. Thiffeault and Boozer \cite{6}, have been further applied Riemannian geometry on the relation between the helicity driven by Riemannian constraints, with the purpose of investigating the onset of kinematic dynamo action. More recently, investigation of the twisted magnetic flux tubes as astrophysical plasmas in Riemannian space has been also performed by Garcia de Andrade, first \cite{7} by building a conformal geometry where the tube is stretching by ideal plasma, where resistivity ${\epsilon}$ vanishes and on a second moment \cite{8}, by investigating the anti-fast dynamo theorem by Vishik \cite{9} as applied to resistive plasmas \cite{8}. This paper is divided into two main parts: In the first it is shown that the sign of helicity of ${\alpha}^{2}$-dynamos, is fundamental on the resulting sign, of constant two-dimensional Riemannian curvature in the dynamo spectrum of ideal plasmas. This result seems to generalize previous result by Chicone et al \cite{10}, by which the Riemannian curvature of the two-dimensional fast dynamos, is always negative. In the second part of the paper, this result is investigated when the dissipation is turn-on, and comparison of these results in undertaken. In the second case, ${\alpha}^{2}$-operator dynamo spectra methods developed by Kirillov et al \cite{11}, the eigenvalue spectrum of the helical filaments orthogonal to Hasimoto soliton surfaces, may indicate the presence, of dynamo action. The consequent investigation of the eigenvalue growth rate of the magnetic fields, leads one naturally to the behaviour of dynamo action. problem of filamentary structures in galactic plasma dynamics have been addressed by Kinney et al \cite{12}, by using the concept of Elss\"{a}sser variables. They also used vortex dynamics structures. In their paper, no dynamo action in plasma has been investigated. Yet more recently Wilkin et al \cite{13} have investigated the existence of dynamo action in turbulent filamentary structures, where small-scale dynamos \cite{14}, can produce these filamentary structures. They showed that, at least in kinematic stage of dynamos, filamentary profiles are preffered, rather than surface structures like ribbons. In their study they made use of the Reynolds magnetic number $R_{m}=\frac{vl}{\epsilon}$, where v and l are respectively, the typical velocities and scales involved in the plasma, while ${\epsilon}$ is the magnetic diffusity. In this paper by making use of ${\alpha}^{2}$-operator dynamo methods developed by Kirillov et al \cite{11}, the eigenvalue spectrum of the helical filaments orthogonal to Hasimoto soliton surfaces \cite{15}, may indicate the presence, of dynamo action. Two kinds of dynamo ${\alpha}$-effect are obtained here in curved plasmas. Both of them have been obtained by Ruediger and Hollerbach \cite{16}, in more restricted manifolds. These are the ${\alpha}=constant$ steady ${\alpha}^{2}$-dynamo on a resistive plasma filament, and the other is the profile ${\alpha}=\frac{\cos{\theta}(s)}{r_{0}}$, where $r_{0}$ is the constant cross-section radius of the curved magnetic flux tube in the thin filament approximation. In this paper, following the Guenther et al approach, one is able to show that a yet simpler eigenvalue analysis can be obtained for ${\alpha}^{2}$-dynamo when one write the induction equation in the Frenet local reference frame, which follows isolated filaments. Thus by assuming the filamentary structure of Wilkin et al, one is led to a dynamo action in three-dimensions. In two-dimensional regime, either the dynamo action cannot be supported or at the best, slow dynamos in plasmas \cite{5} is obtained. Considerable effort to provide simple Riemannian geometrical dynamo models \cite{5}, have been made recently, since the first toy model of a chaotic dynamo on a torus surface, given by Arnold et al \cite{3}. One of the drawbacks of simple models is that in general they do not need to fast dynamo action as appears in solar plasmas and galactic dynamos. Nevertheless maximum possible simplicity, might be an important characteristic to be addressed. Another interesting aspect of the application of differential Riemannian geometry to dynamos is that Anosov two dimensional constant Riemannian curvature spaces has been demonstrated by Chicone et al \cite{10} to be good candidates of kinematic fast dynamos in highly conductive ideal plasmas. Actually, it is shown that the simple Frenet frame used in this paper, the eigenvalue spectra is very similar to the one obtained by Chicone and Latushkin, though main differences between their work and ours, consists in the facts that, they use a dynamo equation which is not a ${\alpha}^{2}$-dynamo; secondly their analysis used a differential forms approach to obtain the dynamo spectrum on surfaces and not filamented dynamos in Frenet frame obtained here. From the mathematical viewpoint, the distinction comes from the fact that here, one uses the differential geometry of curves in three-dimensional Euclidean spaces, rather than the Riemannian geometry of the two-dimensional surfaces in $\textbf{E}^{3}$. In the case of oscillatory dynamos it is interesting to point out that if the helicity changes sign, the curvature does so, thus the the $cos{\theta}$ profile, induces a change in curvature at the equatorial plane of the sun or the Earth. It is important to stress that the helicity considered here is the kinetic helicity and not the magnetic helicity also connected to dynamos, as considered previously by Boozer \cite{17}. Contrary to the kinetic helicity case, the magnetic helicity is necessary for the dynamo action to be supported. The paper is organised as follows: Section II deals with the non-ideal plasmas in Riemannian closed surfaces filled by an ideal dissipative oscillatory ${\alpha}^{2}$-dynamo plasma.  This shows that the constant Riemannian curvature can be positive as well, in order to obtain a fast dynamo action in two dimensional plasmas. Section III deals with the non-ideal plasmas in Riemannian closed surfaces filled by a dissipative oscillatory ${\alpha}^{2}$-dynamo plasma, on filamentary magnetic structures over the surface. In this case slow dynamos are obtained. In the case of $non-{\alpha}$ effect dynamo, such as Chicone et al one, it is shown that a fast dynamo of negative Riemannian curvature can be obtained when the eigenvalues are degenerate. In this section the Frenet frame formalism is applied to two dimensional ${\alpha}^{2}$-dynamos. In previous section Parker cyclonic effect in twisted flux tubes is discussed, with respect to the relations between kinetic helicity and Riemann curvature. Section IV presents future prospects and conclusions.
\newpage
\section{Resonant oscillatory ${\alpha}^{2}$-dynamos in flux tubes}
As pointed out by Zeldovich \cite{1}, ${\alpha}^{2}$-dynamos may exist on a cosmic framework where the ideal plasma resistivity vanishes. More recently, Ruediguer and Hollerbach, have investigated the role of constant and oscilllatory ${\alpha}^{2}$-dynamos, in geodynamos and astrophysical settings. In this section one shall investigate the geometrical and topological structure of magnetic flux tube filled with ideal plasmas, which generate an ${\alpha}^{2}$-dynamo on the Riemannian curvature substrate. Let us then, consider the case of the kinematic ${\alpha}^{2}$-dynamo equation in curved Riemannian two-dimensional manifold. Let us now consider the Riemannian metric describing the twisted magnetic flux tubes as
\begin{equation}
d{{l}}^{2}=dr^{2}+r^{2}d{\theta}^{2}+K(s)^{2}ds^{2}
\label{1}
\end{equation}
where $K(s)^{2}:=[1-{\kappa}(s)rcos{\theta}]^{2}$, r being the radius of the tube cross-section, ${\kappa}(s)=\frac{1}{R(s)}$ the Frenet scalar curvature (R(s) being the external radius of  the tube), and $cos{\theta}(s)$ the description of the oscillatory factor of the twisting of the tube. Here one shall assume that $K(s)\approx{1}$, since in the thin tube approximation, $r\approx{0}$. By substituting this Riemannian line element into the induction dynamo
\begin{equation}
\frac{{\partial}\textbf{B}}{{\partial}t}={\nabla}{\times}({\alpha}\textbf{B})+
{\epsilon}{\Delta}\textbf{B}
\label{2}
\end{equation}
where ${\epsilon}$ is the diffusion constant and ${\Delta}={\nabla}^{2}$ is the Laplacian in the curvilinear Riemannian coordinates. This equation can be considered as the eigenvalue dynamo operator from
\begin{equation}
{\cal{L}}\textbf{B}={\lambda}\textbf{B}
\label{3}
\end{equation}
where the dynamo operator becomes
\begin{equation}
{\cal{L}}={\nabla}{\times}({\alpha})+{\epsilon}{\Delta}
\label{4}
\end{equation}
By applying this dynamo operator, into the magnetic field eigenvetor $\textbf{B}$ as
\begin{equation}
\textbf{B}={B}_{\theta}(r,s)\textbf{e}_{\theta}+B_{s}(r)\textbf{t}
\label{5}
\end{equation}
where due to the physical nature of the magnetic flux tubes the radial component of the magnetic field $B_{r}$ is assumed to vanish. The flow velocity, $\textbf{v}$ is given by
\begin{equation}
\textbf{v}={v}_{\theta}(r,s)\textbf{e}_{\theta}+v_{s}(r)\textbf{t}
\label{6}
\end{equation}
and the divergence-free vector fields of flows and magnetic field obeying the equations
\begin{equation}
{\nabla}.\textbf{v}=0
\label{7}
\end{equation}
The expression ${\nabla}.\textbf{B}=0$ yields
\begin{equation}
{\partial}_{s}B_{\theta}=B_{\theta}{{\kappa}_{0}}^{2}r_{0}sin{\theta}\label{8}
\end{equation}
From the vector analysis formulas 
\begin{equation}
{\nabla}{\times}\textbf{B}=\frac{1}{\sqrt{g}}{\epsilon}^{ijk}[{\partial}_{j}B_{k}]\textbf{e}_{i}
\label{9}
\end{equation}
and the Laplatian operator is
\begin{equation}
{\Delta}\textbf{B}=\frac{1}{\sqrt{g}}{\partial}_{i}[\sqrt{g}g^{ij}{\partial}_{j}]
\label{10}
\end{equation}
where $g=r^{2}$ is the determinant of the Riemann metric of components $g_{ij}$ $(i,j=1,2,3)$ and the Riemannian gradient operator in thin flux tube approximation is 
\begin{equation}
{\nabla}=\textbf{e}_{r}{\partial}_{r}+\frac{1}{r}\textbf{e}_{\theta}{{\partial}}_{\theta}+
\textbf{t}{\partial}_{s}
\label{11}
\end{equation}
Here one has used the following relations between the Frenet frame and the frame $(\textbf{e}_{r},\textbf{e}_{\theta},\textbf{t})$ yields
\begin{equation}
\textbf{e}_{\theta}=-\sin{\theta}\textbf{n}+\cos{\theta}\textbf{b}
\label{12}
\end{equation}
and
\begin{equation}
\textbf{e}_{r}=\cos{\theta}\textbf{n}+\sin{\theta}\textbf{b}
\label{13}
\end{equation}
where one has used the following relations between the Frenet frame and the frame $(\textbf{e}_{r},\textbf{e}_{\theta},\textbf{t})$ yields
One also has made use of the evolution equations of the Frenet frame
\begin{equation}
\frac{d\textbf{t}}{ds}={\kappa}(s)\textbf{n}
\label{14}
\end{equation}
\begin{equation}
\frac{d\textbf{n}}{ds}=-{\kappa}(s)\textbf{t}+{\tau}\textbf{b}
\label{15}
\end{equation}
\begin{equation}
\frac{d\textbf{b}}{ds}=-{\tau}(s)\textbf{n}
\label{16}
\end{equation}
From these expressions one is able to compute the elements of the ${\alpha}^{2}$-dynamo equation. The first term on the RHS of the induction equation is
\begin{equation}
{\nabla}{\times}({\alpha}\textbf{B})=\frac{1}{r_{0}}[(-B_{\theta}{{\kappa}_{0}}^{3}r_{0}sin{\theta}(1+r_{0}))
\textbf{t}+{{\kappa}_{0}}^{2}(B_{s}+\frac{1}{2}B_{\theta}r_{0}{\kappa}_{0}sin{2{\theta}})\textbf{b}]
\label{17}
\end{equation}
 Where one has assumed that the ansatz for the amplification of the magnetic field of the fast dynamo, with exponential growth ${\lambda}$ is $\textbf{B}=\textbf{B}_{0}e^{{\lambda}t}$. Attaching the orthonormal Frenet frame, $(\textbf{t},\textbf{n},\textbf{b})$, to the magnetic axis of the curved and twisted flux tube, the remaining results can be recasted in terms of the Frenet frame expressions of twist, which is ${\partial}_{\theta}=-{{\tau}_{0}}^{-1}{\partial}_{s}$ and curvature scalar ${\kappa}_{0}$. Here, to simplify matters, a helical flux tube, is considered where the constant torsion ${\tau}_{0}$, coincides with the Frenet curvature \cite{18} ${\kappa}_{0}$. Now, one is ready to compute the other RHS of the dynamo equation, or Laplatian operator as
\begin{equation}
{\Delta}={{\partial}_{r}}^{2}+[1-\frac{{{\tau}_{0}}^{2}}{{r_{0}}^{2}}]{{\partial}_{s}}^{2}
\label{18}
\end{equation}
The term on the LHS of the equation is
\begin{equation}
{\partial}_{t}\textbf{B}={\lambda}\textbf{B}+{\gamma}[{\kappa}_{0}(B_{s}-{\kappa}_{0}r_{0}
sin^{2}{\theta})\textbf{n}-B_{\theta}{\kappa}_{0}sin{{\theta}}\textbf{t}+B_{\theta}{{\kappa}_{0}}^{2}r_{0}
cos{{\theta}}\textbf{b}]
\label{19}
\end{equation}
where ${\gamma}=({\omega}_{\theta}-{\omega}_{s}){\kappa}_{0}$. This value shall vanish in the following computations, since one is assuming the resonant mode ${\omega}_{\theta}={\omega}_{s}$ applies between the poloidal ${\omega}_{\theta}$ and toroidal ${\omega}_{s}$ frequencies of the flux tube. Thus by considering the magnetic eigenvector as
\begin{equation}
(B_{\theta},{B_{s}})^{T}
\label{20}
\end{equation}
where T represents the transpose matrix, allows us to express this equation in terms of the local Frenet frame, and to write the three result scalar equations of the kinematic ${\alpha}$ effect dynamo, into the matrix form as
\begin{equation}
\vspace{1mm} $$\displaylines{\pmatrix{{\epsilon}{{\kappa}_{0}}^{3}&{-{\lambda}}\cr
{{\lambda}\cos{\theta}+{\partial}_{s}{\alpha}\sin{\theta}}&{{\epsilon}{{\kappa}_{0}}^{2}}\cr}\cr}$$
\label{21}
\end{equation}
The 2D equation resulting from this is possible since a third equation for the magnetic field components was left out of the matrix. This remaining equation, gives our first eigenvalue as
\begin{equation}
{\lambda}_{1}={\epsilon}{{\kappa}_{0}}^{3}r_{0}\frac{B_{\theta}}{B_{s}}
\label{22}
\end{equation}
This equation shows that this eigenvalue corresponds to a slow dynamo eigendirection. However, as one shall soon see, that the other eigenvalues from the remaining of the spectrum, obtained from the vanishing determinant of the matrix (\ref{21}) is given by
\begin{equation}
cos{\theta}{\lambda}^{2}+\sin{\theta}{\partial}_{s}{\alpha}{\lambda}+{{\kappa}_{0}}^{5}{\epsilon}^{2}=0
\label{23}
\end{equation}
This algebraic second-order equation, yields the remaining part of the spectrum
\begin{equation}
{{\lambda}_{(2,3)}}_{\pm}=\frac{1}{2a}[-b\pm{\sqrt{b^{2}-4ac}}]
\label{24}
\end{equation}
where 
\begin{equation}
a=cos{\theta}\label{25}
\end{equation}
\begin{equation}
b=\sin{\theta}{\partial}_{s}{\alpha}
\label{26}
\end{equation}
\begin{equation}
c={{\kappa}_{0}}^{5}{\epsilon}^{2}
\label{27}
\end{equation} 
Since these eigenvalues obey the following expression
\begin{equation}
lim_{{\epsilon}\rightarrow{0}}{{\lambda}}_{2,3}\ge{0}
\label{28}
\end{equation}
in order to dynamo action be supported (positive sign) or the case of marginal dynamos (equal sign) the limit is
\begin{equation}
{{\lambda}_{(2,3)}}_{\pm}=-{\partial}_{s}{\alpha}tan{\theta}
\label{29}
\end{equation}
Thus to know exactly what is the relation between the kinetic helicity ${\alpha}$, given by
\begin{equation}
{\alpha}=(\textbf{v}.{\nabla}{\times}\textbf{v})=-\frac{cos{\theta}}{r_{0}}
\label{30}
\end{equation}
where $r_{0}$ is the approximately constant radius of the flux tube. Substitution of this $\alpha$ into the equation (\ref{28}) yields
\begin{equation}
{{\lambda}_{(2,3)}}_{\pm}=-\frac{{\kappa}_{0}sin^{2}{\theta}}{r_{0}cos{\theta}}
\label{31}
\end{equation}
Note that in order to support dynamo action, or $\cos{\theta}>0$ and curvature ${\kappa}_{0}<0$, or $\cos{\theta}<0$ and curvature ${\kappa}_{0}>0$. Thus since as has recently been shown \cite{10}, the Riemann tensor is proportional to the Frenet frame scalars curvatures, so one may say that Riemann curvature can be either positive or negative in two-dimensions for dynamo action to be supported. In the first case the domain of ${\theta}$ is $0<Dom{\theta}<\frac{{\pi}}{2}$, while the other domain is $\frac{3{\pi}}{2}<Dom{\theta}<\frac{{\pi}}{2}$. Intersection of both domains implies that one is on the dynamo action of Riemann-flat torus region, or more close to the geometrical axis of the twisted surface. This idea is also in agreement with Arnold et al steady uniform stretching primitive dynamo, with the solely difference that their dynamo is not a ${\alpha}^{2}$-dynamo. The Perm dynamo torus, which is also turbulent, however, Shukurov et al work on dynamo Moebius strip flow, uses dynamo equation without the ${\alpha}$-effect as considered here.

\section{Filamentary steady ${\alpha}^{2}$-dynamos in Riemannian 2D flows}
In this section one presents the main ideas on the spectrum of the induction equation, and subsequent investigation on the possible existence of the ${\alpha}^{2}$-dynamo in $2D$ in magnetic filamentary structures, which gives rise to galactic dynamos. The magnetic lines along the filaments are computed in the Frenet frame $(\textbf{t},\textbf{n},\textbf{b})$, where the tangent vector $\textbf{t}$, is along these lines, while the respectively, normal and binormal vectors $\textbf{n}$ and $\textbf{b}$ belong to an orthogonal plane to the magnetic filament. This frame vectors obey the following evolution equations
Here ${\kappa}$ and ${\tau}$ are Frenet curvature and torsion scalars. Let us assume that, the flow $\textbf{v}=v_{b}\textbf{b}$,where $v_{b}={\kappa}_{0}$, lays over a solitonic Riemannian Hasimoto surface \cite{15}. Note that, though one considers here that the modulus of the flow $v_{b}$ is constant, the flow is not necessarily laminar due to the dynamical unsteady nature of the frame vector $\textbf{t}$. This equation shall be expanded below, along the Frenet frame as
\begin{equation}
\textbf{B}(s,t)=B_{s}(s,t)\textbf{t}(t,s)+B_{n}\textbf{n}+B_{b}\textbf{b}\label{32}
\end{equation}
\newpage
Proceeding in the way analogous to the last section, one obtains the spectrum
\begin{equation}\vspace{1mm} $$\displaylines{\pmatrix{{\lambda}-{\alpha}{{\kappa}_{0}}&{\epsilon}+{\partial}_{s}{\alpha}\cr
{{\partial}_{s}{\alpha}}&{{\lambda}}\cr}\cr}$$
\label{33}
\end{equation}
The divergence-free magnetic vector equation
\begin{equation}
{\nabla}.\textbf{B}={\partial}_{s}B_{s}-{\kappa}_{0}B_{n}=0 \label{34}
\end{equation}
and the dynamo operator above, one is able to write the eigenvalue equation
\begin{equation}
det[{\lambda}\textbf{I}-{\cal{L}}_{\epsilon}]=0\label{35}
\end{equation}
As in section II, the matrix of eigenvalues leads to the third-order algebraic equation
\begin{equation}
{\lambda}^{2}-{{\kappa}_{0}}{\alpha}{\lambda}+({\epsilon}+{\partial}_{s}{\alpha})=0
\label{36}
\end{equation}
Here ${\alpha}=<\textbf{v}.{\nabla}{\times}\textbf{v}>$ is the helicity of ${\alpha}^{2}$-dynamo, which is given by
\begin{equation}
{\alpha}=-{{\kappa}_{0}}^{2}\label{37}
\end{equation}
Thus 
\begin{equation}
{\partial}_{s}{\alpha})=0
\label{38}
\end{equation}
which simplifies equation (\ref{36}) to
\begin{equation}
{\lambda}^{2}-{{\kappa}_{0}}{\alpha}{\lambda}=0
\label{39}
\end{equation}
which has the simple solution
\begin{equation}
{\lambda}=-{{\kappa}_{0}}^{3}
\label{40}
\end{equation}
Thus the curvature ${\kappa}_{0}<0$. Thus since this result represents a degenerate eigenvalue both eigendirections of the surface would have a negative curvature. Therefore, the Riemannian or Gaussian curvature is positive since Gaussian curvature is the product of the filament curvatures over the surface.

\section{Filamentary structures in 3D kinematic dynamo spectra}
Let us now consider the magnetic kinematic dynamo, which considers the regular induction equation in 3D with non-zero plasma resistivity ${\epsilon}$, to compare with the two dynamo cases considered in the previous sections. The induction equation is
\begin{equation}
\frac{{\partial}\textbf{B}}{{\partial}t}={\nabla}{\times}[\textbf{v}{\times}\textbf{B}]+
{\epsilon}{\Delta}\textbf{B}\label{41}
\end{equation}
where ${\Delta}={\nabla}^{2}$ is the Laplacian operator. Here we also assume that the same decomposition as above is done, with the difference that now the binormal component of the magnetic field $B_{b}$ does not vanish. Besides here the magnetic helicity does not appear in the induction equation. In this case the divergence free law or the absence of magnetic monopole remains the same. By considering the rescaling $v_{0}:=1$, the three scalar induction equation obtained from the decomposition of the vector induction equation along the filaments is, along $\textbf{t}$, $\textbf{n}$ and $\textbf{b}$ directions, are
\begin{equation}
[{\lambda}-{{\kappa}_{0}}^{2}{\epsilon}]B_{s}-{\kappa}_{0}B_{n}+
{\epsilon}{{\kappa}_{0}}^{2}B_{b}=0\label{42}
\end{equation}
\begin{equation}
{\kappa}_{0}B_{s}+[{\lambda}+2{{\kappa}_{0}}^{2}{\epsilon}]B_{n}-2{\kappa}_{0}B_{b}=0\label{43}
\end{equation}
\begin{equation}
-{\alpha}{{\kappa}_{0}}^{2}B_{s}-2{\kappa}_{0}B_{n}+({\lambda}+{\epsilon}{{\kappa}_{0}}^{2})
=0\label{44}
\end{equation}
In this 3D dynamo, the eigenvalue spectrum equation
\begin{equation}
det[{\lambda}\textbf{I}-{D}_{\epsilon}]=0\label{45}
\end{equation}
where now the dynamo operator matrix $D_{\epsilon}$, can be written as
\begin{equation}
$$\displaylines{\pmatrix{{-{\epsilon}{{\kappa}_{0}}^{2}}&{-{\kappa}_{0}}&
{{\epsilon}{{\kappa}_{0}}^{2}}\cr{{\kappa}_0}&{2{\epsilon}{{\kappa}_{0}}^{2}}&
{-2{\kappa}_{0}}\cr{-{\epsilon}{{\kappa}_{0}}^{2}}
&{-2{\kappa}_{0}}&{{\epsilon}{{\kappa}_{0}}^{2}}\cr}\cr}$$
\label{46}
\end{equation}
As in section II, the matrix of eigenvalues leads to the third-order algebraic equation
\begin{equation}
{\lambda}^{3}-{\epsilon}{{\kappa}_{0}}^{2}{\lambda}^{2}-{{\kappa}_{0}}^{2}
[1+\frac{{\epsilon}}{2}{{\kappa}_{0}}^{2}]{\lambda}+2[1-\frac{1}{2}
[1-4{{\kappa}_{0}}{\epsilon}]]
{{\kappa}_{0}}^{2}=0
\label{47}
\end{equation}
Though in general, complete third-order algebraic equations and higher, are very complicated one shall address here a special case of physical interest to dynamo theory. In this case the equation is reduced to a second-order equation. In this case, one shall consider that the growth rate of the magnetic field ${\lambda}$ is very small on a kind of slow dynamo. Thus, the third-order term in ${\lambda}$, could be truncated, or neglected  and the polynomial equation (\ref{27}) would be reduced to
\begin{equation}
-{\epsilon}{{\kappa}_{0}}^{2}{\lambda}^{2}-{{\kappa}_{0}}^{2}
[1+\frac{{\epsilon}}{2}{{\kappa}_{0}}^{2}]{\lambda}+2[1-\frac{1}{2}
[1-4{{\kappa}_{0}}{\epsilon}]]
{{\kappa}_{0}}^{2}=0
\label{48}
\end{equation}
A simple particular solution of this equation can be obtained as
\begin{equation}
{\lambda}=2^{\frac{1}{3}}[1-\frac{1}{2}(1-4{{\kappa}_{0}}{\epsilon})]
{{\kappa}_{0}}^{\frac{2}{3}}\label{49}
\end{equation}
which in the limit of ideal plasmas where resistivity ${\epsilon}$ vanishes, one obtains
\begin{equation}
lim_{{\epsilon}\rightarrow{0}}{\cal{R}}{\lambda}=[{\frac{{{\kappa}_{0}}}{2}}]^{\frac{2}{3}}
\label{50}
\end{equation}
Here, ${\cal{R}}{\lambda}$, is the real part of ${\lambda}$ which in general has complex roots, indicating that the dynamos oscillates. Thus since this limit is positive, and does not vanish (slow dynamo), a fast dynamo solution is obtained from this dynamo spectrum. In the above computations the incompressible flows.
\newpage
are compatible with the above definition of the flow. Just for comparison, one shall reproduce here the Riemannian three dimensional spectrum obtained by Chicone et al as
\begin{equation}
$$\displaylines{\pmatrix{{-{\epsilon}}&{0}&
{0}\cr{0}&{-{\epsilon}}&
{-{\kappa}_{0}}\cr{{\epsilon}{{\kappa}_{0}}^{2}}
&{1-{\kappa}_{0}{\epsilon}}&{-{\epsilon}{{\kappa}_{0}}^{2}}\cr}\cr}$$
\label{51}
\end{equation}
One may note that these matrices are very similar in character and dependence on the magnetic diffusivity constant and constant curvature. Chicone et al did not considered in detail this three dimensional dynamo and no mention has been done on turbulent ${\alpha}^{2}$-dynamo.
\section{Conclusions}
 By making use of mathematical tools from operator spectral theory, an investigation of the spectra of two kind of dynamos is performed. The first is the ${\alpha}^{2}$-dynamo, so useful in turbulence and geodynamos. The second is the chaotic dynamo so useful in MHD dynamo plasma theory. A new fast dynamo solution comes out from this spectrum investigation, where the eigenvalue spectrum is obtained from a particular solution of the third-order algebraic equation. A more complete panorama of the present solution can be obtained by performing the graphic between the growth rate of the magnetic field and the Reynolds magnetic number $Rm$. Since the Rm is the inverse of the parameter of diffusion ${\epsilon}$ certainly the fast dynamo presented here implies a high Rm, which by the relation obtained by Wilkin et al \cite{2}, for the filaments thickness $l_{\epsilon}=l_{0}{Rm}^{-\frac{1}{2}}$, one may conclude that the fast dynamo obtained in the last section is actually a thin filament dynamo with thickness which depend upon curvature as $l_{\epsilon}=2{{\kappa}_{0}}^{\frac{1}{2}}l_{0}$, which shows that the filament Since curvature is related to the folding, one may say that a faster dynamo can be obtained in three-dimensions with the enhancement of folding, what does not happen in two-dimensions. An important part of the paper deals with turbulent dynamos in Riemannian twisted magnetic flux tubes, which is a kind of generalization of dynamo Moebius strip flow. recently addressed by Shukurov et al. Resonance process in oscillatory dynamos is important to design possible new experiments in Perm dynamo torus facility.\section{Acknowledgements}
 Several discussions with Dmitry Sokolov are highly appreciated. I also thank Andrew Soward for kindly sending me a reprint of his work on slow dynamos. Financial  supports from UERJ and CNPq are gratefully acknowledged.
 
  \end{document}